# Picosecond laser induced hybrid groove structures on Ti6Al4V bio-alloy to accelerate osseointegration

*Sunita Kedia [1, 3], Rahul Checker [2, 3], Santosh Kumar Sandur [2, 3] and J. Padma Nilaya [1, 3]

[1]Laser & Plasma Technology Division, Bhabha Atomic Research Centre, Mumbai 400085
[2]Bio-Organic Division, Bhabha Atomic Research Centre, Mumbai 400085
[3]Homi Bhabha National Institute, Training School Complex, Anushaktinagar, Mumbai 400094

*Corresponding author: skedia@barc.gov.in

**Abstract:** Regulating cell growth, extracellular matrix deposition and mineralization of artificial implants are some important parameters that decide the longevity of implants in the body. Picosecond laser induced hybrid groove structures have been shown to improve these properties of the Ti6Al4V bio-alloy. Two hybrid structures containing groove patterns with periodic and non-periodic substructures therein were generated on Ti6Al4V by varying the extent of laser pulse overlapping on sample surface. Laser induced alteration in surface topography, chemical composition and wettability of Ti6Al4V resulted in 3-fold increase in the rate of hydroxyapatite growth, 2.5-fold increment in protein adsorption and 2-fold enhancement in cell adhesion in comparison to pristine sample. While the periodic substructure was found to guide cell growth, the non periodic sub structure offered homogenous growth leading to higher overall cell density on the substrate surface.

1. Introduction

Extensive research has clearly revealed the superiority of titanium based alloy (Ti6Al4V) over other metalic bio-alloys e.g., stainless steel and cromium-cobolt alloy for orthopedic and dental applications because of its higher strength to weight ratio, corroison resistance and biocompatibility [1-3]. However, Ti6Al4V being bio-inert, its bonding with surrounding tissue is weak and develops over a prolonged period of time leading to poor and slow osseointegration and prosthesis loosening of titanium based implants in the long run. The osseous wound healing is a complex process comprising of four dominant phases viz; haemostasis phase, inflammatory phase, proliferative phase and remodelling phase that evolve progressively in time. Immediately following the surgery, haemostasis phase begins with the recovery of the wound that last over few miniutes to an hour and involves adsorption of blood plasma proteins on implant surface. Thereafter, the immune system is activated and this process, termed as inflammatory phase, happens over hours to days. Proliferative phase involves formation of new extra cellular matrix, development of woven bone and its subsequent mineralization by hydroxyapatite (HAP). This process may continue for weeks. In the final phase of remodelling, which begins after 3 weeks of surgery and can last for years, the woven bone get replaced by lamellar (newly formed) bones along the load oriented direction [4]. The rate of all these processes critically decides the quality of initial fixation and longevity of artificial implants in human body. This process, commonly known as osseointegration, represents bonding of artificial implant with living bone structurally and functionally and therefore, proper osseointegration is necessary as its deficiency can lead to formation of fibrous tissues and osteolysis [5]. Needless to say, the physiochemical properties of the implant surface play an important role in this process. Modifications of the surface of an existing implant can further aid its integration with body e.g., augmention of surface contact area by patterning or texturing increases the residence time and contact area of proteins, ions, minerals, and cells on biomaterial surface increasing the cytocompalitibility, differentiation, protein adhesion, osseointegration effectively in comparison to smoother surface of a pristine sample. Several techniques such as blasting, texturing, lithography,and chemical etching have been explored to alter the surface behaviour of biomaterials [6]. However, blasting and chemical etching lead to uncontrolled patterening and surface contamination [7] while lithography is a mask based laborious process involving many steps [8]. Laser induced surface patterning/texturing is not only devoid of any of these disadvantages but has fascinated researchers with its unique advantages of remote, single step, non-contact and fast processing capabilities.

Laser surface texturing (LST), utilising short and ultrashort pulse duration lasers as tools to create such modifications, has attracted the attention of research fraternity of medical field [9-11]. While peripherial distortion i.e heat affected zone is unaviodable in the usage of short pulse (nanosecond) lasers [12], the high cost, high maintenance and long down time restricts commercialization of ultrashort-ultrafast (femtosecond) lasers. A prudent tradeoff is working with a picosecond laser (ps-laser) which is a robust system, can be used in ambient conditions, is easy to handle and is easily available, relatively compact and requires low maintenance.Various macro/micro/nano topographies generated on Ti6Al4V and SS using ps-laser are known to produce antiadhesive and antifouling surfaces with applications in food and medical industries [13,14]. Yu *et al.* used laser to texture Ti6Al4V to promote cell adhesion, proliferation and growth of new blood vessels [15], while Krzywicka *et al.* textured Ti6Al7Nb bio-alloy to correlate change in surface energy of the sample with microbial adhesion [16]. Thus, it is beyond doubt that judicious modification of the implant surface can be used as a means to successfully fine-tune its interaction mechanism with body and accelarate osseointegration.

To gauge the effect of surface treatment on the functionality of a biomaterial under in-vitro conditions, biomimetic deposition of bone apatite in simulated body fluid (SBF) has been widely analysed. Cell adhesion test has proven to be a valuable tool to evaluate the possibility of induced cytotoxicity, if any, due to laser treatment. Protein adhesion, that plays a crucial role in cellular activities including attachment, proliferation and differentiation prior to cell adhesion, is also an important parameter that decides the suitability of a material as a bio-implant. In this work, we report, substantial improvement in the biofunctionality of Ti6Al4V alloy by picosecond laser induced hybrid groove structuring that simultaneously enhanced its ability to support and augment HAP growth, boost protein adsorption and cell adhesion, essential prerequisites for bio-integration, as compared to pristine samples. Here, the emission of a picosecond frequency doubled Nd:YAG laser was suitably focused onto the surface of the bio-alloy to create two distinct groove patterns one with a periodic substructure and another with a non-periodic substructure by changing the sample scan speed, that in turn, changed the overlap and consequently the morphology of the structures formed on the surface. Both the laser treated surfaces have been found to exhibit a significant improvement in various aspects of biofunctionality when compared to the pristine. Further, the periodic substructure supported a guided cell growth while the non-periodic substructure allowed homogenous growth and hence a higher overall cell density.

## 2. Materials and Methods

### 2.1 **Materials**:

Commercially procured Ti6Al4V sheet of thickness ~1 mm was cut into (10 x 10) mm$^2$ pieces. Top surface of the cut sample was polished using 200, 800 and 1200 grit papers in that order and was followed by ultrasonic cleaning in acetone, ethanol and water each for 10 min. SBF was prepared from reagents procured from SRL that are listed in Table-1a. For protein adsorption and cell growth tests, DMEM, MTT (3-(4,5-Dimethyl-2-thiazolyl)-2,5-diphenyl-2H-tetrazolium bromide), and dimethyl sulfoxide (DMSO) were procured from M/S. Sigma Chemical Co. (USA). Foetal bovine serum (FBS) was purchased from GIBCO BRL. Mouse normal fibroblast cells (L929) were purchased from European Health Protection Agency, UK.

### 2.2 **Methods**:

#### *2.2.1 Laser Surface Texturing:*

Surface texturing of Ti6Al4V was performed by employing an Nd-YAG laser (Model no: PL2250, Ekspla make) capable of emitting ~ 15mJ per pulse of 30 ps duration at 532 nm at a reptition rate of 10Hz. The laser beam was focused on to the sample to a spot of diameter ~30 µm using a 10 cm focal length lens. Two groove structrues with distinctly different microstructrues therein, termed as hybrid structures, were generated by translating the sample at a speed of 50 µm/sec in one case and 5 µm/sec in the other using a computer controlled translational stage. The laser power was fixed at 5 mW. A vertical shift of more than 40 µm after each horizontal scan ensured that there was no overlap of the laser spot along the y-axis. The hybrid structure formed for higher sample scan speed condition contained grooves with periodic ripple structure therein and the sample was termed as G-P. In the second case where the scan speed was relatively lower, random structures were generated inside the groove and the sample was named as G-NP. This difference in the microstructure, understandably, was due to different overlapping factor of laser spots on the sample surface in the two cases. The overlapping factor can be estimated using equation-1 [17]

$$Pulse\ overlap = 1 - \frac{v}{b \times f} \qquad (1)$$

where v is sample scan speed, b is width of focal spot and f is laser repetition rate. At the two different scan rates employed viz., 50 µm/sec, 5 µm/sec, the pulse overlap factor was ~83% and 98% for G-P and G-NP, respectively.

### 2.2.2 Growth of bone like apatite:

Time dependent biomimetic of HAP was observed by immersing samples in SBF which was prepared in-house by following a standard protocol [18]. Briefly, in order to prepare 1 L SBF, the reagents in accordance with their corresponding weight as listed in Table-1a were dissolved in a mixture of water (960 mL) and 1M HCl (40mL) completely employing a magnetic stirrer. The SBF so prepared by this method mimics the blood plasma in terms of ion concentration to a large extent as can be seen in Table-1b. Especially, concentrations of Ca and P, that mainly participate in the process of HAP growth, match well with that in blood plasma. Four samples of each type, were prepared, immersed in100 mL of SBF individually (Fig. 1c) and placed in an incubator at 37°C for a known period such as 1 day, 3days, 15days, and 25 days after which the samples were removed dried in air and characterised.

Table-I: (a) List of reagents and their corresponding weights used for the preparation of SBF, (b) Nominal ion concentration in SBF and in human blood plasma [18], and (c) pictorial representation of samples immersed in SBF

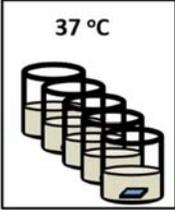

### 2.2.3 Characterization:

The change in surface topography of the sample after laser treatment and HAP growth was analysed using table top mini scanning electron microscope (SEM) (M/s. S&C). A 3D optical profilometer was used to estimate the average surface roughness ($S_a$) of the sample. The wettability of the sample was estimated by water contact angle measurement. Micro-Raman analysis and X-ray diffraction techniques were used to confirm the growth of HAP. The chemical composition and Ca/P ratio of the grown layer on samples when immersed in SBF was studied using Energy Dispersive X-ray spectroscopy (EDS).

### 2.2.4 Protein adsorption:

The effect of altered surface topography on protein adsorption capacity of the surface was studied by soaking the samples in 300 µg/mL fetal bovine serum (FBS) solution (Himedia, India) at 37 °C and 5% $CO_2$ for one hour. At the end of the incubation time, they were rinsed twice with PBS (Phosphate buffered solution) to remove softly adhering proteins and the amount of non-adsorbed proteins was quantified through the Bradford assay (BIO-RAD Protein Assay, BIO-RAD, USA). The percentage of firmly adsorbed serum proteins on samples was then calculated.

### 2.2.5 Cytotoxicity and Adhesion of primary cells:

Cytotoxicity and adhesion of primary cells of the LST samples was estimated by MTT assay method [19]. For cell culture, L929 (mouse normal fibroblast cells), from European Health Protection Agency, UK were used. Cells were cultured in DMEM medium supplemented with 10% fetal bovine serum and antibiotics (100 U/ml-penicillin, 100 µg/ml-streptomycin) at 37 °C in 5% $CO_2$ atmosphere and sub-cultured at 80% confluence. The L929 cells (10,000 cells/well in 2 ml complete media) were seeded in flat-bottom 24-well plates and allowed to grow for 72 hrs after which 200µl MTT reagent (5 mg/ml) was added to each well and these cells were incubated on P, G-P and G-NP samples at 37 °C for 4 hr. The medium was discarded and formazan crystals were solubilized by adding DMSO (200 µl per well) and absorbance was measured at 570 nm.

### 2.2.6 Immunofluorescence microscopy:

L929 cells (0.4 x $10^6$) were seeded in 6 well plates in the presence of samples. After 48 h incubation, cells were fixed with 4% paraformaldehyde solution. The cells were permeabilized using 0.5% Triton X-100 (5 min), washed with PBS and blocked using 5 % BSA (1 h). Cells were then stained with tubulin antibody overnight, washed three times with PBS and incubated with FITC labeled secondary antibody (1:200) at room temperature for 1 h. Each sample was washed three times with PBS, nuclei were stained with propidium iodide (50µg/ml) and cells were visualized using IncuCyte S3 Imaging System (Sartorius, Germany) [20].

**Statistical Analysis:** Statistical significance was assessed by Student's "t"-test. Data points represent mean from three replicates and two independent experiments were performed.

## 3. Results

### 3.1 Topography analysis:

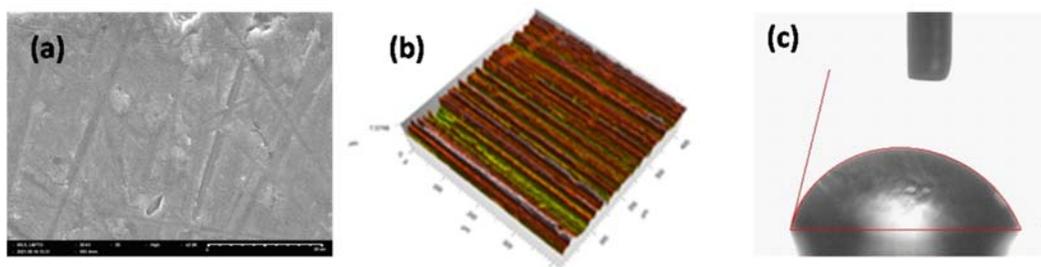

**Fig. 1: (a) SEM image, (b) optical image, and (c) Photograph of water droplet on P**

The SEM images in Fig. 1a show the surface morphology of P sample after polishing and cleaning. The $S_a$ of the sample, as estimated from the optical profilometer image (Fig. 1b) was ~0.87 μm and the sample was found to be hydrophilic in nature with water contact angle ~ 78°, as shown in Fig. 1c.

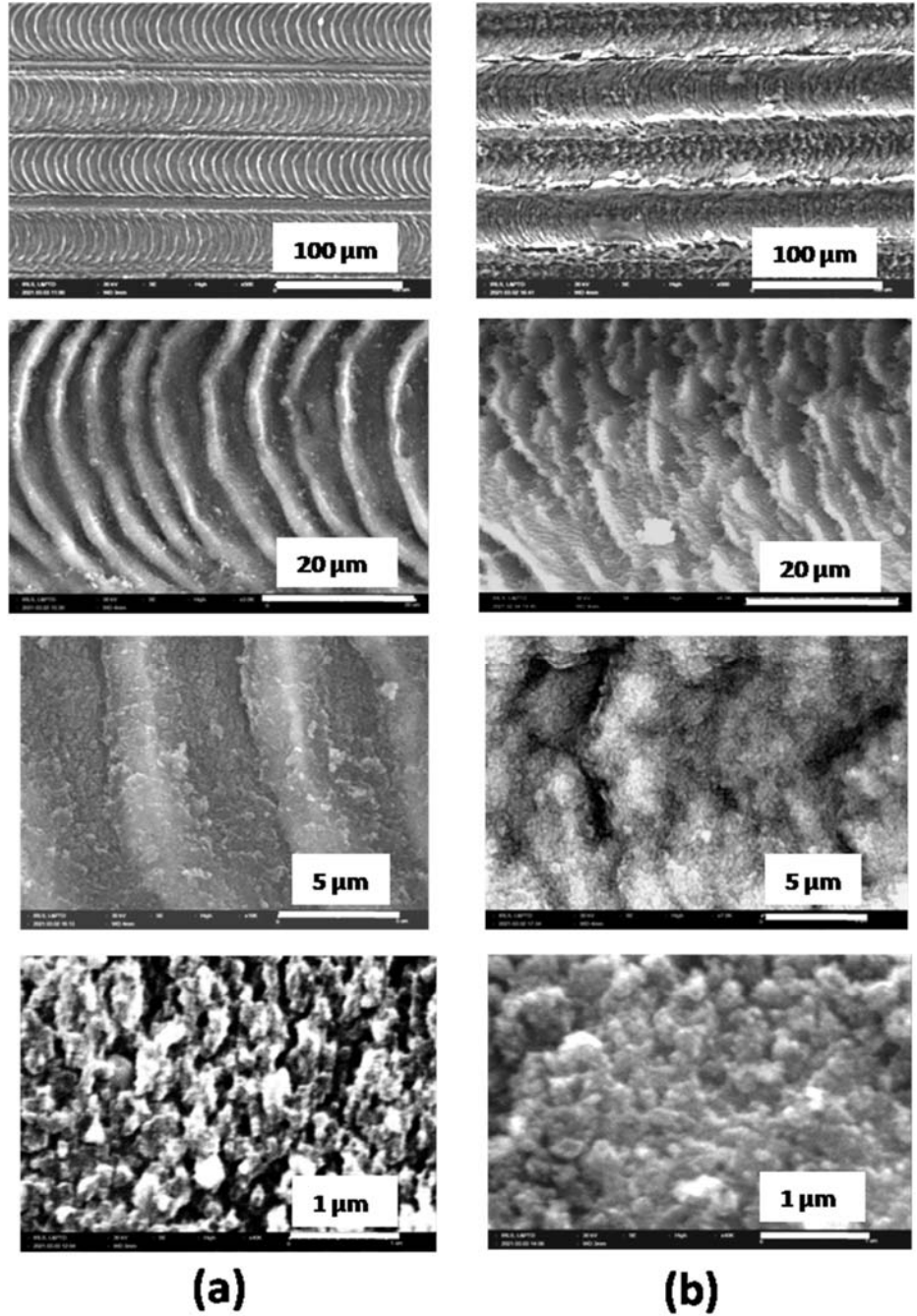

**Fig. 2: SEM image of (a) G-P and (b) G-NP at different magnifications**

Multiple parallel grooves with significantly distinct sub-structure therein were observed in SEM images of G-P and G-NP (Figs. 2a and 2b) respectively. In case of G-P, well arranged periodic ripple structure with average separetion of ~5 μm and porous microstructures within the ripples can be seen in SEM images shown in different scales of magnification. However, in case of G-NP, the grooves are finer with relatively denser microstructures. Although, the laser power was maintained constant in these two cases, such distinct variation in the microstruture can be attributed to different percentage of pulse overlapping. As stated before, increase in scan speed (50 μm/sec) reduces the pulse overlap to 83% in case of G-P while it is 98% in G-NP (5 μm/sec). While a repetition rate of 10 Hz is too low to cause an increase in temperature due to accumulated heat, the repeated cycles of melting and resolidification due to higher overlap seen by G-NP gives rise to the denser microstructures.

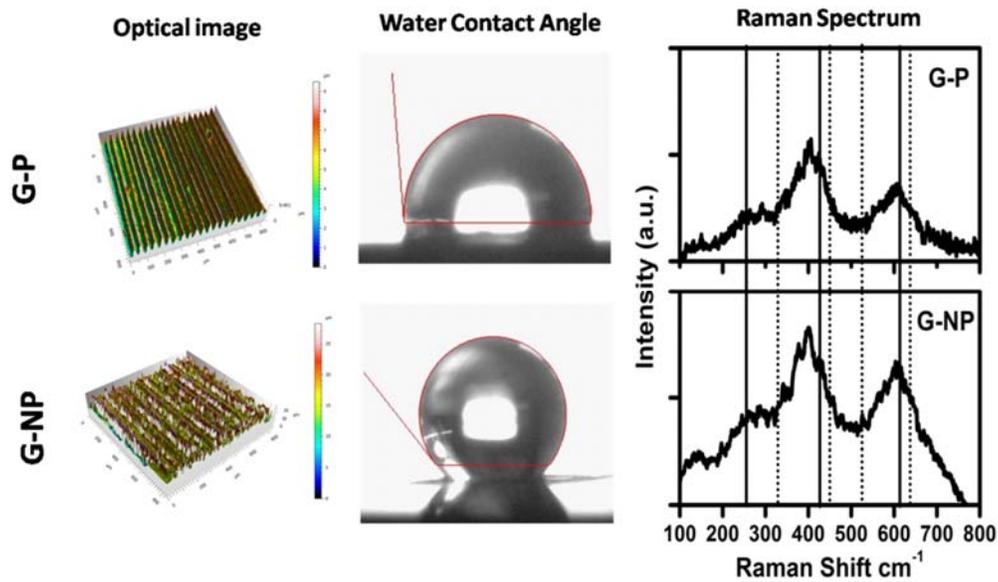

**Fig. 3: 3D profilometer image, water contact angle and Raman spectra (solid and dotted lines indicated wavenumber corresponding to anatase and rutile phases) of G-P and G-NP**

Fig. 3 shows the optical profilometer images, photograph of water droplet and micro-Raman analysis of G-P and G-NP samples. The $S_a$ for G-P and G-NP were 1.40 μm and 3.04 μm, both being higher for the laser treated samples compared to pristine. The tendency towards increased hydrophobicity of the treated samples was established with measured water contact anglesof 94 degrees and 129 degrees respectively. The Raman analysis revealed the generation of both anatase (wavenumber: 143 $cm^{-1}$, 395 $cm^{-1}$ and 515 $cm^{-1}$ ; solid lines in Fig. 3) and rutile phases (wavenumber: 231 $cm^{-1}$, 438 $cm^{-1}$ and 604 $cm^{-1}$ dotted lines in Fig.3) of $TiO_2$ on G-P and G-NP. These details are summarised in Table-II for easy reference.

*Table-II: Experimental parameters and data of surface analysis of pristine and laser treated samples.*

| Sample | P | G-P | G-NP |
|---|---|---|---|
| Sample scanning speed | As received sample (P) | 50 µm/sec G-P | 5 µm/sec G-NP |
| Groove width (µm) | - | 50 | 45 |
| Pulse overlapping | - | 83% | 98% |
| $S_a$ (µm) | 0.87 | 1.40 | 3.04 |
| WCA | 78º | 94º | 129º |

### 3.2 Growth of bone like apatite

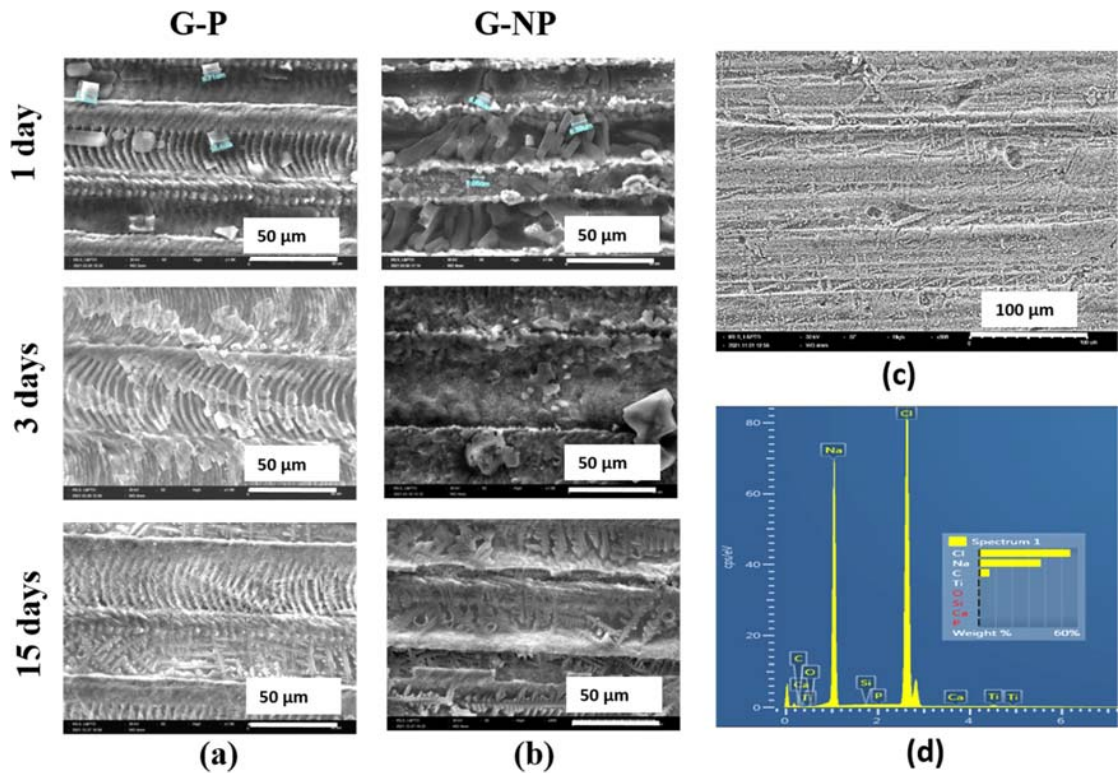

**Fig.4: Time dependent growth of bone like apatite on (a) G-P, (b) G-NP, (c) after 15 days on P and (d) Typical EDS spectrum of apatite grew on samples**

Fig. 4 shows the SEM images of G-P and G-NP samples immersed in SBF for 1, 3 and 5 number of days. Within 1 day of immersion, discrete nucleation giving rise to cube like structures with average size of ~10 µm were observed on G-P, and ~30 µm long cuboids with relatively larger density were found on G-NP surface. The second set of samples was removed after 3 days from SBF and long chains of connected cubic structures were observed on G-P, while filled grooves were seen on G-NP surface. We note here that, similar crystal deposition on the surface of Ti6Al4V was also observed by Procher *et al.* [21]. The set of sample removed after 15 days showed filled grooves with dendrite like structures in both the cases. Similar dendrite growth was observed on P sample after 15 days of immersion, as shown in Fig. 4c. The EDS analysis of these samples indicated that the crystals were rich in sodium and chloride and a typical image of EDS analysis of one of the sample is shown in Fig. 4c. This clearly points to the fact that until 15 days no signature of apatite growth was detected on the surface of prisitne or laser treated samples.

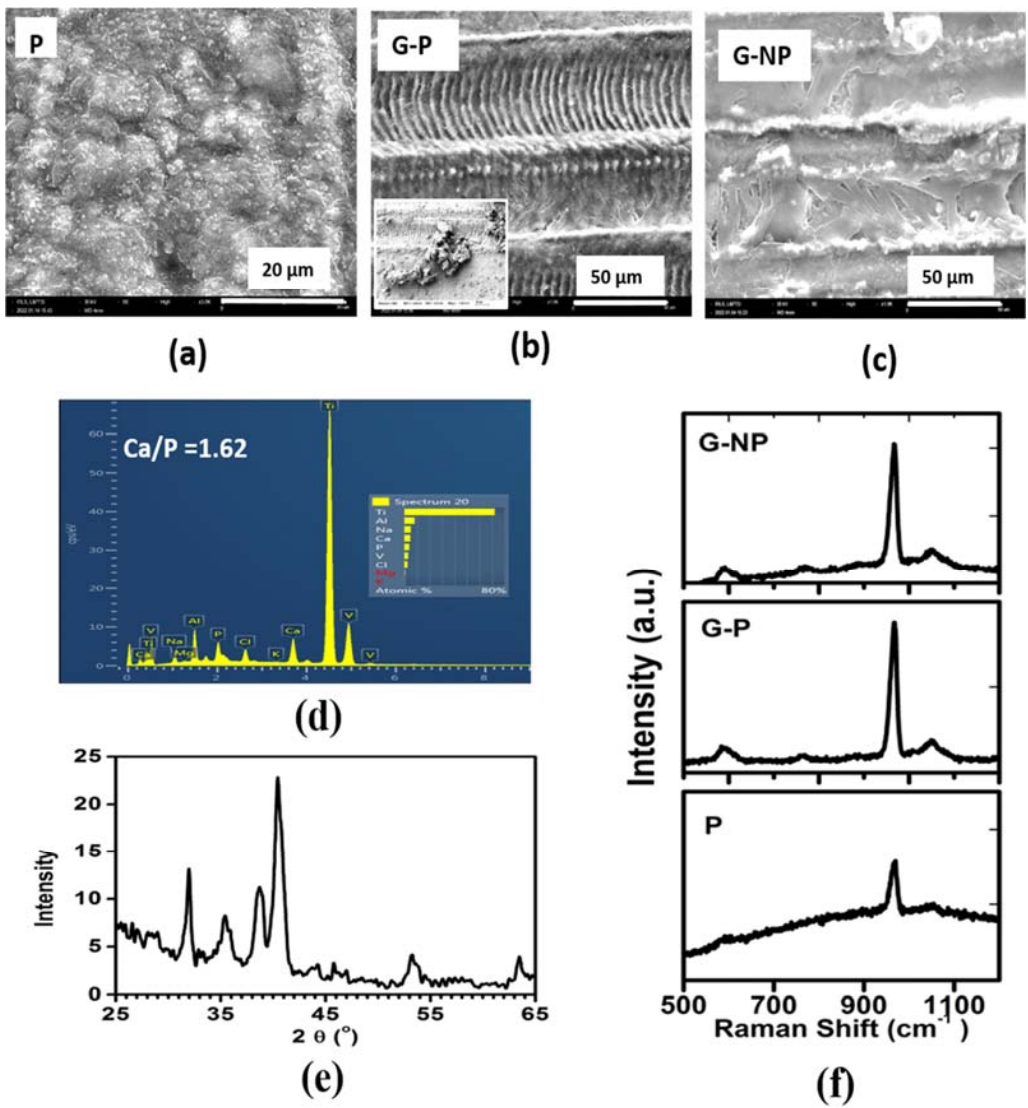

**Fig. 5: Growth of HAP on after 25 days(a)P, (b) G-P, (c) G-NP, (d) EDS analysis, (e) GI-XRD plot and (f)micro-Raman spectrum of HAP grew on P, G-P, and G-NP samples**

Fig 5a-5c are the SEM images of P, G-P and G-NP samples when immersed in SBF for 25 days, respectively. Discrete nucleation of HAP on P sample and filled grooves on G-P and G-NP was clearly observed. The magnified image of HAP on G-P is shown in inset of Fig. 5b. The EDS analysis of HAP revealed a healthy Ca/P ratio of ~1.62 on G-P (Fig. 5d) and ~1.60 on G-NP and ~1.51 for P samples (not shown here). The GI-XRD peaks at 26º and 32º as depicted in Fig 5e further confirmed the growth of HAP (JCPDS 896438) on G-P. Fig. 5f is the micro-Raman analysis of HAP on S-P, G-P and G-NP. The characteristic Raman peak of HAP at 960 cm$^{-1}$ attributed to internal modes of the $PO_4^{3-}$ tetrahedral is seen for all the three samples. Along with 3-fold intense characteristic peak of HAP, two peaks at 589 cm$^{-1}$ and 1046 cm$^{-1}$ assigned to O-P-O bending and asymmetric P-O stretching, were also observed on laser treated samples [22]. Multiple Raman peaks with highly intense characteristic peak indicates faster and superior quality growth of HAP on laser treated samples.

*3.3 Protein adsorption & Cell growth*

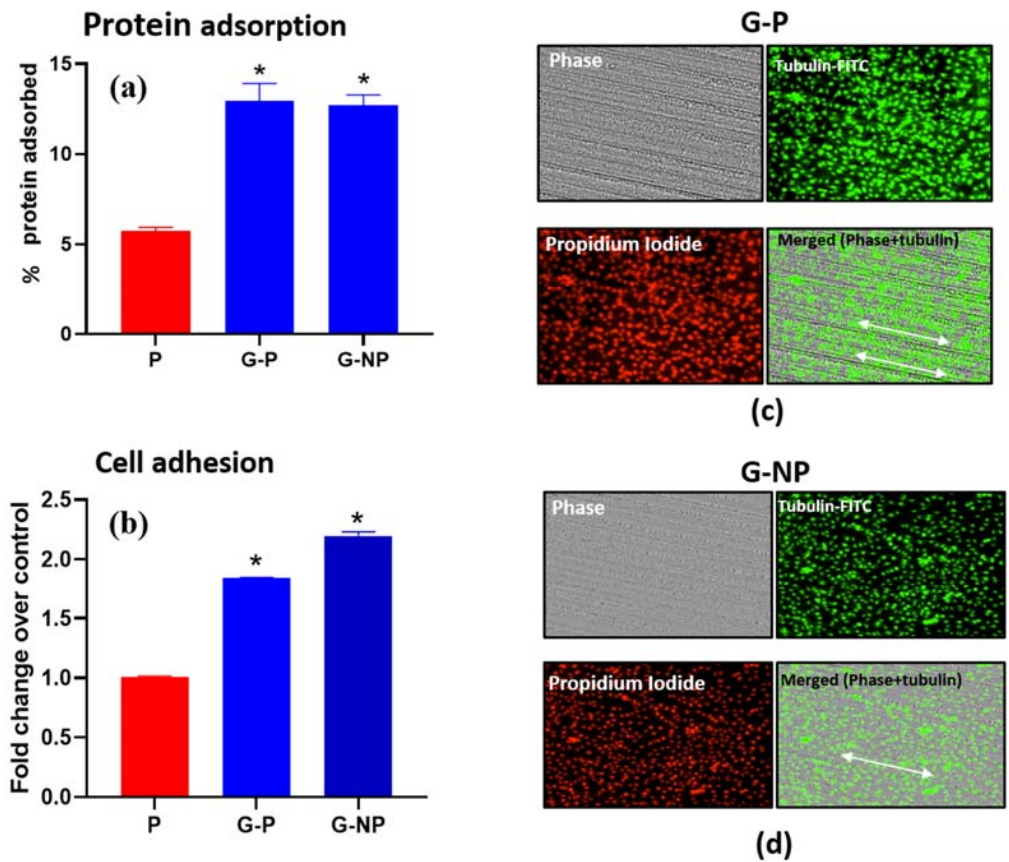

**Fig. 6: (a) Percentage of protein adsorption and (b) Cell adhesion and (c & d) microscopic image of cells on sample G-P and G-NP respectively**

Experiments were carried out to understand the effect of altered surface topography on protein adsorption capacity of the surface. The adsorption of serum protein improved by ~ 2.5 fold on the surface of G-P and G-NP samples in comparison to P sample (Fig. 6a) which indicated increased affinity of Ti6Al4V towards serum proteins post laser treatment. MTT assay and fluorescence microscopy were done to understand the growth kinetics and morphology of cells on different samples. As shown in Fig. 6b, a marked difference in the growth kinetics of L929 cells on G-P and G-NP samples as compare to P sample is observable. The data indicated a significant enhancement in the number density of cells ( ~2 fold) on laser treated samples as compared to P sample. Among the laser treated samples, the cell density was found to be slightly more on G-NP sample. Further, the florescence microscopy images of L929 fibroblast cells clearly show that cells appear to grow at a higher density along the groove (marked with white arrow in the figure) on G-P sample (Fig. 6c). On the other hand, cells adhere and grow at random sites on G-NP as can be observed in Fig. 6d. Thus, the periodic sub-structures in the groove of G-P supported regular arrangement of cells along the groove.

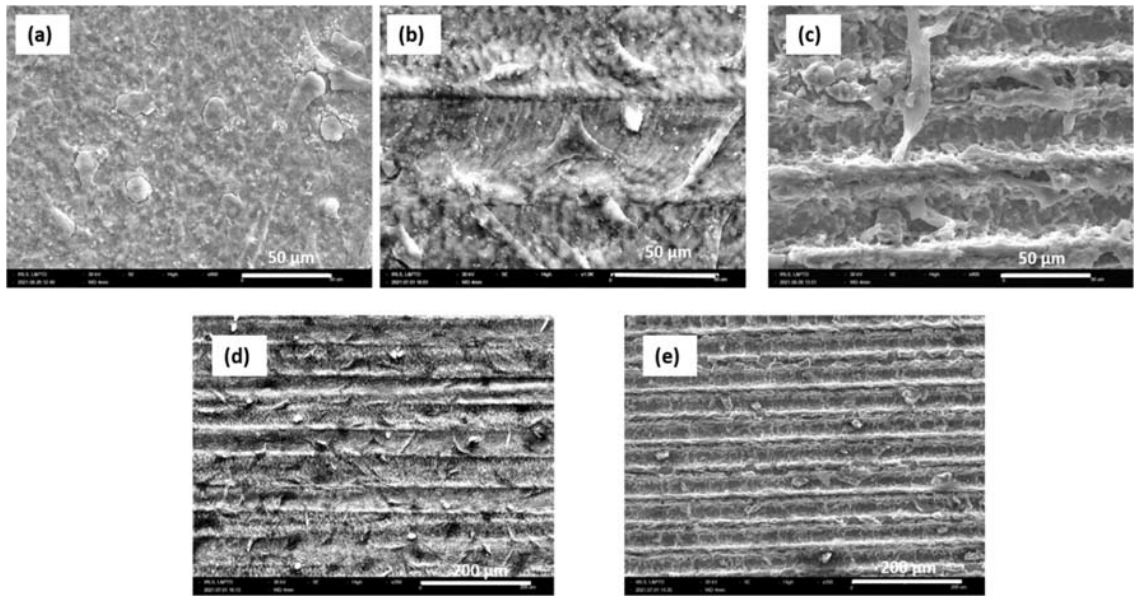

**Fig. 7: SEM image of cells on (a) P, (b), (d) G-P and (c), (e) G-NP sample surface**

SEM images shown in Fig. 7 were acquired to see the morphology of L292 cells on the surface of the sample after 4 hr of incubation. Most of the cells were observed in circular morphology on P sample (Fig. 7a), on contrary, cells grew to larger dimensions asuming tubular, triangular and polygonal shapes on G-P (Fig. 7b) and G-NP (Fig. 7c) samples. A number of filopodia developing in different directions and connected to the substrate were seen on grooved surfaces. The cytoskeletons of most of the cells with different shapes can be seen lying within the grooves and filopodia attached to the groove wall. Polygonal cells with higher number of filopodia are observed on periodic grooved sample surface indicating well spreading of cells. These features of filopodial extensions and higher number density indicate superior cytocompatibility of the modified surfaces in comaprison to P sample. To appreciate it better, SEM images of larger areas of the sample surface are shown in Fig. 7d and 7e, respectively.

4. Discussion

Effect of surface topography on osseointegration can be an arduous and extensive study as a multitude of patterns can be created on a substrate surface with sub features ranging from micron to nanometer sizes and many times a combination of both [23-24] simply by changing the experimental conditions and operating parameters of the laser. Achieving guided cell growth, extracellular matrix deposition and faster mineralization of artificial implant through surface modification can be an important aspect for osseointegration and bone implant longevity. Laser induced groove patterns have been investigated in the past and shown to possess the capability to support cell alignment, directional migration and contact guidance [17, 25-26]. Here, laser modification technique has been employed to create groove patterns of two different types on Ti6Al4V bio-alloy and in-vitro biological tests such as, growth of HAP, protein adsorption and cell adhesion have been performed. Surface characterization showed a general increase in surface roughness, decreased in surface wettabilty and generation of mixed (anatase and rutile) phases of titanium-oxide on the surface of Ti6Al4V post laser treatment. The surface features generated due to laser treatment are found to be strongly dependent on the overlap factor here, other parameters being maintained identical. In vitro bio-tests clearly show the superiority of LST samples that is discussed in detail below. To be noted here that all these biological effects come into action simultaneously in case of a bio implant placed in a body. However, to understand the effect of laser surface treatment independently on each of these aspects we have resorted to separate in vitro testing.

*4.1 Apatite like growth:*

HAP ($Ca_{10}(OH)_2(PO_4)_6$) is calcium orthophosphate salt with Ca/P molar ratio of about 1.667 and contains hydroxide groups ($OH^-$). During biomimetic in SBF, the ability the surface to support HAP build-up on it is a measure of the bioactivity of the material. A biomaterial that shows good bioactivity in SBF will readily allow growth of HAP on its surface, support adhesion, differentiation and multiplication of cells on its surface with HAP layer working as a natural scaffold when it comes in contact with blood plasma[21].

During initial days of immersion, distinct deposition of crystals with various surface morphologies was observed on G-P and G-NP. Chains of cubic structures on G-P may be because of the periodic hybrid sub-structures in the groove of the sample which provided regular sites for deposition. Growth of HAP was found on all samples only after 25 days of immersion, however growth was superior on laser treated grooved surfaces in general as confirmed from EDS, GI-XRD and Raman analysis. It is known that the hydrophilic nature of a surface can support growth of HAP [27]. Contrary to this, we observed that the growth of HAP was better on hydrophobic surface under similar deposition conditions. This may be due to two important attributes the surfaces acquire as a result of laser treatment, viz., surface roughness and oxide layer. The laser induced microstructures increased surface roughness and hence improve interaction and dwell time of ions in SBF on the sample surface. Additionally, the grooved surface contains mixed phase of anatase and rutile and it is known that rutile phase plays an important role on apatite growth as the lattice structures of rutile and apatite matches[28].Among groove samples, the Ca/P ratio and Raman characteristic peak of HAP was higher for G-P sample, indicating the groove structure with periodic sub-structure supported HAP growth superiorly. Since the distribution of anatase and rutile phases on the surface of laser treated samples may be different and inhomogeneous, the exact effect of these phases on the growth of HAP was not be estimated.

### 4.2 Protein adsorption:

The protein binding capacity of a surface improves with increasing surface area[29] and the role of protein in promoting cell adhesion and hence bio-compatibilityof the material is well documented[30]. Ionic interaction, chemical bonding and wettability of a sample are the factors that decide adsorption of protein [31]. In this work, we observed a 2.5 fold increment in percentage of protein adsorbedon laser treated grooved surfaces. According to Yoshida *et al.*the electrostatic interaction between protein and titanium is the dominating factor for such improvment [32]. Due to higher surface area of sample, the wetting anisotropy of the grooves by which a droplet would stick to the edges of the grooves and stretch along the groove long axis where convex edges are better accessed by proteins [33] and the increased hydrophobic character post laser treatment that attracts the hydrophobic patches in protein [34] may all be responsible to varying degrees for the observed phenomenon of improved protein adsorption capability of laser treated grooved samples.

*4.3 Cell adhesion:*

Cell proliferation, adherence and migration are the primary factors affecting implant suitability and operation success rate. Interestingly, we observed that laser modification could enhance the adhesion and growth of normal fibroblasts by ~2 fold. This may be attributed to the fact that the groove width and sub-micron structures that provided additional anchoring sites to cells helped them adhere better leading to a better growth kinetics. A significant variation was observed in the morphology of the cells adhered to G-P and G-NP. The proliferation was in definite directions when periodic anchoring sites were available. Hence density of cells was found to be more along the grooves in case of G-P. However, such alignment was not seen in G-NP, indicating that the periodic ripple structure inside the grooves manipulated the cells to adhere in a particular manner. Then aperiodic hybrid structure in G-NP provided random anchor points for the proliferation of the cells, which in turn, we believe is responsible for the higher number density observed here. Thus, it is not only the microstructuring but also the type of sub-structure within these microstructures that further adds to the improving the functionality of bio-alloy.

5. **Conclusion**

In conclusion, this study reveals that Ti6Al4V sample surfaces with picosecond laser induced hybrid groove structrues exhibited a superior growth of HAP, adhesion of bovine serum and L929 normal cells leading to an improved osseointegration and biocompatibility as against pristine samples. The altered surface properties due to laser treatment e.g., topography, wettability and oxide states, increased surface roughness etc were responsible for such improvement. Further, it has been also demonstrated that the cell proliferation depends on the type of hybrid structure on the substrate, e.g., periodic sub-structure guides the growth along the grooves.

**Ethical statement:**

This is to declare that, the submitted work is original and is not published elsewhere and the manuscript is not submitted elsewhere for publication. The order of authors and the corresponding author are correct in the submission.

**Acknowledgment:** Authors acknowledge Dr. A. K. Sahu, G&AMD, BARC for EDS analysis and P. S Gaikward, IRLS, BARC for technical help.

# Reference

[1] K. Prasad, O. Bazaka, M. Chua, M. Rochford, L. Fedrick, J. Spoor, R. Symes, M. Tieppo, C. Collins, A. Cao, D. Markwell, K. K. Ostrikov, and K. Bazaka, Materials 31, 884 (2017).

[2] M. Niinomi and M. Nakai, Int. J. Biomater. 2011, 836587 (2011).

[3] M. Kaur and K. Singh, Materials Science and Engineering C 102, 844 (2019).

[4] H. Terheyden, N. P. Lang, S. Bierbaum and B. Stadlinger, Clin. Oral. Impl. Res.,1(2011).

[5] T. Goto, Clin Calcium 24, 265 (2014).

[6] S. Mukharjee, S. Dhara and P. Saha, Jornal of Manufacturing Processes 65, 119 (2021).

[7] P. Schupbach, R. Glauser and S. Bauer, Hindawi International Journal of Biomaterials 1, (2019)

[8] N. Idota, T. Tsukahara, K. Sato, T. Okano, and T. Kitamori, Biomaterial 30, 2095 (2009).

[9] Y. Xu, W. Liu, G. Zhang, Z. Li, H. Hu, C. Wang, X. Zeng, S. Zhao, Y. Zhang and T. Ren, Journal of the mechanical behaviour of biomedical materials 109, 103823 (2020)

[10] Z. Yu, G. Yang, W. Zhang and J. Hu, Journal of Materials Processing Tech. 255, 129 (2018).

[11] M. M. Calderon, M. M. Silvan, A. Rodriguez, M. G. Aranzadi, J. P. G. Ruiz, S. M. Olaizola and R. J. M. Palma, Sci. Rep 6, 36296 (2016).

[12] S. Kedia, S. Shaikh, A. G. Majumdar, M. Subramanian, A. K. Sahu and S. Sinha, Adv. Matt. Lett. 10, 825 (2019).

[13] F. H. Rajab, C. M. Liauw, P. S. Benson, L. Li and K. A. Whitehead, Colloids and Surface B: Biointerfaces 160, 688 (2017).

[14] F. H. Rajab, C. M. Liauw, P. S. Benson, L. Li and K. A. Whitehead, Food and Bioproducts Processing 109,29 (2018).

[15] Z. Yu, S. Yin, W. Zhang, X. Jiang and J. Hu, J Biomed Mater. Res. , 1 (2019).

[16] M. Krzywicka, J. Szymanska, S. Tofil, A. Malm and A. Grzegorczyk. J. Funct. Biomater. 13, 26 (2022).

[17] S. Kedia, S. K. Bonagani, A. G. Majumdar, V. Kain, M. Subramanian, N. Maiti and J. P. Nilaya, Colloid and Interface Science Communications 42, 100419 (2021).

[18] T. Kokubo and H. Takadama, Biomaterials 27, 2907 (2006).

[19] R. Checker, D. Pal, R. S. Patwardhan, B. Basu, D. Sharma, and S. K. Sandur, Free Radical Biology and Medicine 143, 560 (2019).

[20] B. Singh, R. S. Patwardhan, S. Jayakumar, D. Sharma, S. K. Sandur, Journal of Photochemistry & Photobiology, B: Biology 213, 112080 (2020).

[21] P. Prochor and Z. A. Mierzejewska, Materials 14, 2059 (2021).

[22] C. S. Ciobanu, F. Massuyeau, L. V. Constantin, and D. Predoi, Nanoscale Research Letter 6, 613 (2011).

[23] C. Liang, H. Wang, J. Yang, Y. Cai, X. Hu, Y. Yang, B. Li, H. Li, H. Li, C. Li and X. Yang, ACS Appl. Mater. Interfaces 5, 8179 (2013).

[24] B. E. J. Lee, H. Exir, A. Weck, and K. Grandfield, Appl. Surf. Sci. 441, 1034 (2018).

[25] O. Raimbault, S. Benayoun, K. Anselme, C. Mauclair, T. Bourgade, A. M. Kietzig, P. L. G. Lauriault, S. Valette and C. Donnet, Materials Science and Engineering C 69, 311 (2016).

[26] C. Leclech and C. Villard, Frontiers in Bioengineering and Biotechnology 8, 551505 (2020).

[27] W. Boonrawd, K. R. Awad, V. Varanasi and E. I. Meletis, Surf. Coat. Technol. 414, 1 (2021).

[28] J. J. Jasinki, Mater. Proc. 2, 8 (2020)

[29] M. S. K. Zaqout, T. Sumizawa, H. Igisu, T. Higashi, and T. Myojo, J. Occup. Health 53, 75 (2011).

[30] P. Lagonegro, G. Trevisi, L. Nasi, and L. Parisi, Dent. Mater. J. 37, 278 (2018).

[31] Y. Kusakawa, E. Yoshida and T. Hayakawa, BioMed Research International (2018).

[32] E. Yoshida and T. Hayakawa, Dental Materials Journal 32, 883 (2013).

[33] P. Elter, R. Lange and U. Beck, Langmuir 27, 8767 (2011)

[34] R. R. Behera, A. K. Das, A. K. Das, A. F. Hasan, D. Pamu, L. M. Paddey and M. R. Sankar, Journal of Alloys and Compounds 842, 155683 (2020).